\documentclass[prl, showpacs,twocolumn]{revtex4}
\usepackage{bm}
\usepackage{graphicx}%
\begin{document}
\title
{Magnus force and acoustic Stewart-Tolman effect in type II
superconductors}
\author{V.\,D.\,Fil$^1$, D.\,V.\,Fil$^{2}$, A.\,N.\,Zholobenko$^1$,
N.\,G.\,Burma$^1$,
 Yu.\,A.\,Avramenko$^1$,
J.\,D.\,Kim$^3$, S.\,M.\,Choi$^3$, and S.\,I.\,Lee$^{3,4}$}
\affiliation{%
$^1$B.\,Verkin Institute for Low Temperature Physics and
Engineering, National Academy of Sciences of Ukraine, Lenin av. 47
Kharkov 61103, Ukraine\\ $^2$Institute for Single Crystals,
National Academy of Sciences of Ukraine, Lenin av. 60, Kharkov
61001, Ukraine\\ $^3$Pohang University of Science and Technology,
Pohang, 794784, Korea
\\
$^4$Korea Basic Science Institute, Daejeon, 305333, Korea}
\begin{abstract}
At zero magnetic field we have observed an electromagnetic
radiation from superconductors subjected by a transverse elastic
wave. This radiation has an inertial origin, and  is a
manifestation of the acoustic Stewart-Tolman effect. The effect is
used for implementing a method of measurement of an effective
Magnus force in type II superconductors. The method does not
require the flux flow regime and allows to investigate this force
for almost the whole range of the existence of the mixed state. We
have studied  behavior of the gyroscopic force in nonmagnetic
borocarbides and Nb. It is found that in borocarbides the sign of
the gyroscopic force in the mixed state is the same as in the
normal state, and its value (counted for one vortex of unit
length) has only a weak dependence on the magnetic field. In Nb
the change of sign of the gyroscopic force under the transition
from the normal to the mixed state is observed.
\end{abstract}


\maketitle

The Stewart-Tolman (ST) effect, the emergence of an electrical
current in a coil under an accelerate or decelerate rotation, is a
famous experimental proof for the electron nature of conductivity
of metals.  The inertial force should give a contribution to the
electrical currents excited by a transverse acoustic wave
propagating in the metal (the acoustic ST effect). Nevertheless,
it was believed that this effect is negligible and does not have
any practical use. As far as we know, up to now there was no
efforts to register inertial force in acoustic experiments. In our
study we have observed the acoustic ST effect and use it for a
development of a powerful method of investigation of the
gyroscopic forces acting on a moving vortex in type II
superconductors.

Our study was motivated by the present situation in understanding
of the nature of the gyroscopic force (the effective Magnus
force\cite{son}) in superconductors. Most of all, it concerns the
so called Hall anomaly - the change of sign of the Hall voltage at
the superconducting transition. According to current theoretical
conceptions, this effect may have  both the microscopic origin,
connected with peculiarities of the electron spectrum \cite{2},
and the macroscopic one: the appearance of the transverse force
under large (much large than the core size) displacements of the
vortices  in the pinning potential \cite{2n}. The flux flow
experiments did not allow to separate these two contributions. The
macroscopic effects may remove the genuine (microscopic) Hall
anomaly or, on the contrary, mimic it. For instance, in Nb the
Hall anomaly was observed or not, depending on the sample quality
\cite{3n}.

To overcome this difficulty we provide measurements of the Magnus
force under an oscillatory motion of the vortices with a small (of
the atomic scale) amplitude of the displacements. It allows to
exclude the influence of the pinning forces. Another advantage of
the method is the ability to fulfill measurements in almost the
whole range of the existence of the mixed state.

\begin{figure}
\includegraphics[width=8cm]{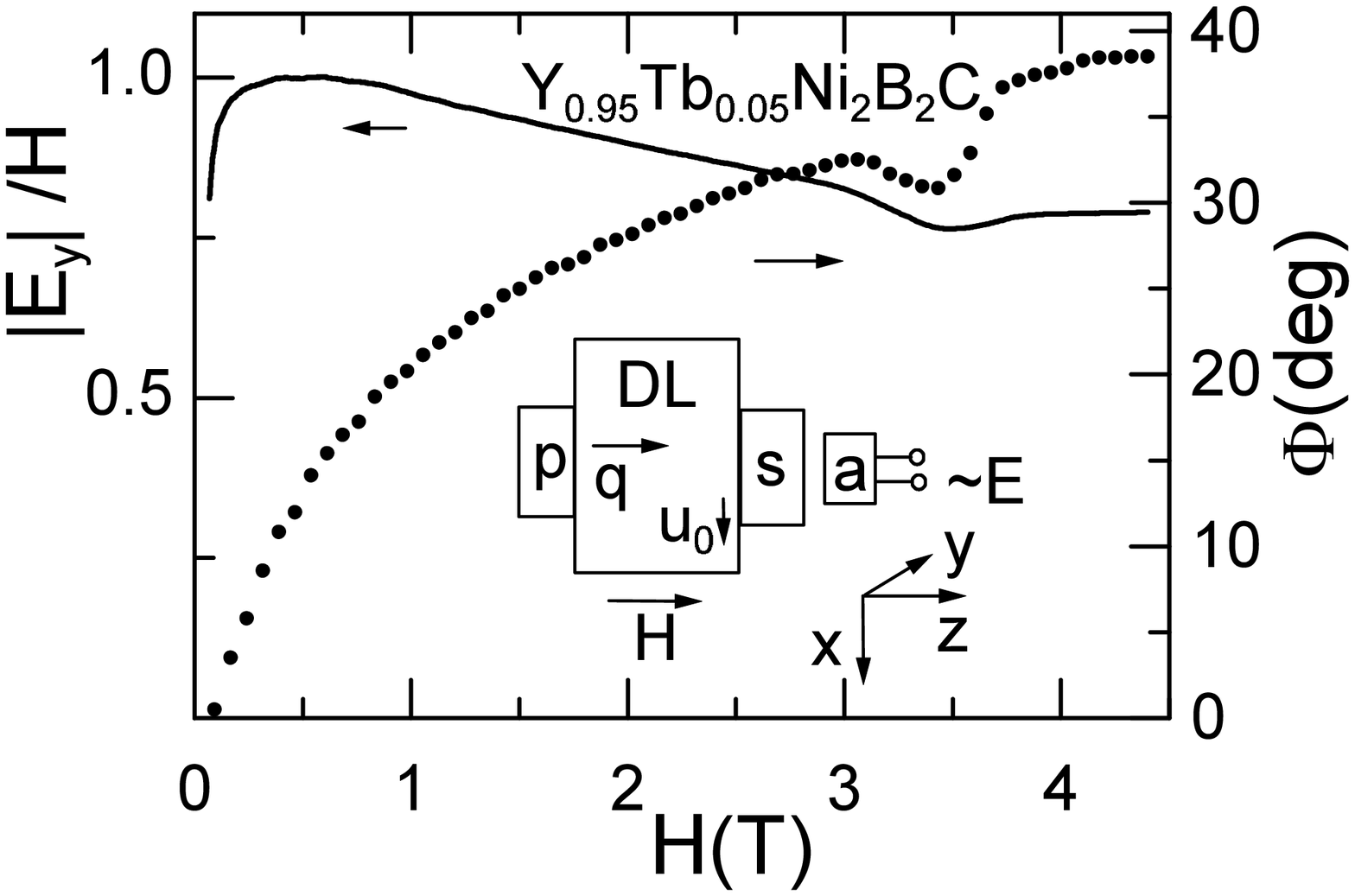}
 \caption{\label{f1} The modulus (solid line) and the phase (dots) of $E_y(H)$ in Y$_{0.95}$Tb$_{0.05}$Ni$_2$B$_2$C
 at $T=1.7$ K ($H_{c2}=3.8$ T). The ratio $|E_y|/H$ is normalized as described in the text. Inset -
 the sketch of the experiment (p, the piezotransducer, DL, the delay line, s, the sample, and a, the antenna)}
\end{figure}

The method is based on the measurements of the amplitude and the
phase of an electromagnetic field (EF) radiated from the
superconductor due to the vortex oscillations excited by a
transverse elastic wave. The scheme of the experiment and the
frame of reference are shown in Fig.\,\ref{f1}(inset).
Technically, the method is close to the one used in \cite{3}. A
key new point is the registration not only large $E_y$, but also
small $E_x$ component of the EF field. The $E_x$ component in the
Meissner state is caused solely by the acoustic ST effect.

In our study we used the single crystals of non-magnetic
borocarbides (YNi$_2$B$_2$C, Y$_{0.95}$Tb$_{0.05}$Ni$_2$B$_2$C and
LuNi$_2$B$_2$C) and Nb as the objects of the investigations. The
borocarbide single crystals were prepared by the same method as in
\cite{4}. The samples have the shape of platelets with the
thickness $\sim 0.5$ mm and the transverse size $\sim 3$ mm. In
all samples the  C$_4$ axis was oriented orthogonal to the
platelet plane, which was used as the radiated surface.  The
working frequencies are 54$\div$55 MHz, and the intensity of the
exciting signal is $\sim$10 W/cm$^2$. The details of the measuring
technics are given in \cite{5}.

In the normal state the $E_x$ and $E_y$ components  can be found
from the joint solution of the Maxwell equations and the kinetic
equation. With accounting for the gyroscopic forces this problem
was solved in \cite{6}. Let the vector of the elastic
displacements is aligned along the $x$ axis (${\bf u}(z,t)=(u_0
\cos (q z) e^{i\omega t}, 0,0)$). In the local limit ($q l\ll 1$,
$q$ is the wave number, and $l$ is the mean free path) and for
$|\Omega| \tau \ll 1$ ($\Omega$ is the cyclotron frequency,  and
$\tau$ is the relaxation time) the electrical field radiated from
the sample at $z=0$ has the form:
\begin{eqnarray}\label{4a}
  &&E_{x}^{(n)}=[1-i \beta_d] X u_{ST} + [1-X] X   \Omega\tau
    u_{ind}, \\
\label{5a}
  &&E_{y}^{(n)}=X u_{ind},
\end{eqnarray}
where $X=k^2_n/(1+k^2_n)$ with the  dimensionless parameter
$k_n^2=4\pi i\omega \sigma_0/q^2 c^2$ ($\sigma_0=ne^2\tau/m$, the
static conductivity), $u_{ST}$ and $u_{ind}$ are the extraneous
forces: $u_{ST}=m_e\omega^2 u_0/|e|$ is the ST inertial force
($m_e$ and $e$ are the mass and the charge of free electron,
respectively) and $u_{ind}=i\omega H u_0/c$ is the inductive
force, $H$ is the external magnetic field. The sign of charge of
the carriers is included into the definition of $\Omega$. The term
in Eq.(\ref{4a}) $\propto i\beta_d$ is caused by the deformation
interaction. For the quadratic spectrum of the carriers
$\beta_d=(1/5)(v_F/s) ql$, where $v_F$ is the Fermi velocity, and
$s$ is the sound velocity. For the magnetic field used in the
experiment $|u_{ind}/u_{ST}|=\Omega/\omega\gg 1$. This strong
inequality makes the magnitude of the second (gyroscopic) term in
Eq. (\ref{4a}) comparable or larger than that of the ST term, and
allows to neglect the term $\propto u_{ST}$ in Eq. (\ref{5a}).

In the mixed state  the EF can be evaluated from the following
system of equations
\begin{eqnarray}\label{1} &&
 \frac{d^2 {\bf E}}{d z^2}=\frac{4 \pi i\omega}{c^2} {\bf j},\\
\label{2} && {\bf j}=\sigma\left( {\bf E} +{\bf u}_{ST}
  +\frac{i\omega}{c}[{\bf w \times B}]\right),\\
\label{3}  && \frac{1}{c}[{\bf j \times
B}]+(i\omega\eta+\alpha_L)({\bf u}-{\bf
  w})\cr && +i\omega\alpha_M [({\bf u}-{\bf w})\times \frac{{\bf B}}{|B|}]=0,
\end{eqnarray}
where $B$ is the magnetic induction, ${\bf j}$ is the current,
${\bf w}$, the displacement of the vortex lattice, $\sigma=n_s
e^2/i\omega m$, the a.c. conductivity in the superconducting state
($n_s$ is the superfluid density), the dynamical parameters
$\eta$, $\alpha_L$, $\alpha_M$, are the viscosity, the Labush
parameter, and  the Magnus coefficient, correspondingly. Eq.
(\ref{1}) is the Maxwell equation, Eq. (\ref{2}) is the matter
equation that connects the current with the field ${\bf E}$ and
the extraneous forces, and Eq. (\ref{3}) is the Galilean invariant
equation of motion of the vortex lattice, normalized to the unit
volume \cite{7}. The boundary condition for the system
(\ref{1})-(\ref{3}) is $d{\bf E}/d z|_{z=0}\approx 0$ \cite{6}.

Taking into account that for typical superconductors $\alpha_M\ll
\eta$ and $q^2 \lambda_L^2\ll 1$ ($\lambda_L$ is the London
penetration depth)  we obtain
\begin{eqnarray}\label{4}
 &&E_{x}^{(m)}=[1-i \beta_{dm}(B)] X(B)u_{ST}+\cr &&[1-X(B)] X(B)  \left[\frac{i\omega
  \alpha_M}{i\omega\eta+\alpha_L}\right] u_{ind},\\
\label{5} &&  E_{y}^{(m)}=X(B)  u_{ind}.
\end{eqnarray}
Here $u_{ind}=i\omega B u_0/c$,  and $X(B)=k^2_m/(1+k^2_m)$ with
$k_m^2=4\pi(i\omega\eta+\alpha_L)/(q^2 B^2)$. The deformation
interaction is not included into Eqs. (\ref{1})-(\ref{3}). But
since the structure of Eqs. (\ref{4}), ({\ref{5}) is similar to
one of Eqs. (\ref{4a}), ({\ref{5a}), we add the deformation
correction into Eq. (\ref{4}) phenomenologically. One can assume
that the quantity $\beta_{dm}$ is approximated as
$\beta_{dm}(B)\approx \beta_d B/H_{c2}$. Going ahead, we emphasize
that for the measurement of $\alpha_M$ we do not need the exact
form of $\beta_{dm}(B)$

The experimental procedure consists in the measuring of the
relative changes of the fields $E_x$ and $E_y$ under a variation
of $H$. To obtain the dynamical parameters from the experimental
data one should set the reference points for the amplitude and the
phase of $E_x$ and $E_y$. In other words, at certain $H$ we should
determine the relation between $E_x$, $E_y$ and $u_0$.

According to the  Bardin-Stephen estimate the viscosity
$\eta\propto B$. Therefore, at H close to $H_{c1}$ the quantity
$|k_m^2|\gg 1$, and the modulus and the phase of $X$ approaches to
1 and 0, correspondingly. Let us count the phase of $E_y$ from its
value at $H\to H_{c1}$. Then $\arg E_y=\arg X $. On the other
hand, at $H=H_{c2}$ the quantity $|X(H_{c2})|=[1+\tan^2 arg
X(H_{c2})]^{-1/2}$ and it is convenient to choose such units for
$|E_y|$ that satisfy the relation
$|E_y(H_{c2})|/H_{c2}=|X(H_{c2})|$.

If the Ginzburg-Landau parameter $\kappa\gg 1$, one can neglect
the difference between $B$ and $H$ in almost the whole range of
$H$ where the mixed state exists. Then, at $H\sim (5\div 10)
H_{c1}$ (in which  $|k_m^2|\gg 1$) the quantity
$|E_y(H)|/H=|X(H)|$ (in the units of $|E_y|$ chosen) should
approach  unity. The fulfillment of this condition can be
considered as an independent test for the measuring procedure
proposed. If this condition fails it can be due to a strong
non-uniformity of the conductivity near the surface of the sample
(see \cite{3}), and in this case one cannot use the relations
(\ref{4}), (\ref{5}) for the analysis of the results of the
measurements. In the experiments presented in this paper the above
mentioned condition was fulfilled with the accuracy $\sim 5\%$.
The typical example of the experimental data is presented in
Fig.\,\ref{f1}. From these data one can find the dependencies
$\eta(H)$ and $\alpha_L(H)$ \cite{8}.

At $H=0$ and $T\ll T_c$ the deformation correction  is frozen,
and, as follows from Eqs. (\ref{1}), (\ref{2}), $E_x=u_{ST}$. It
means that under such conditions the inertial force becomes the
only source for the EF radiated from the sample. Therefore, it is
convenient to choose the amplitude and the phase of the $E_x$
signal at $H=0$ and $T\ll T_c$ as the reference points for the
complex quantity $E_x(H,T)$.

At $T>T_c$ and $H=0$ the component $E_x$ is given by the first
term in the r.h.s. of Eq. (\ref{4a}). In this case it contains the
contribution of the deformation force. But in rather dirty samples
($|k_n^{-2}|>\beta_d$) the $E_x$ field is determined in the main
part by the ST effect. In such samples  the amplitude of the
radiated EF should increase, while its phase should decrease under
transition to the superconducting state. The experimental
dependencies $|E_x(T)|$ and $\arg E_x(T)$ presented in
Fig.\,\ref{f2} demonstrate  such
 behavior.  We consider these results as the first
experimental observation of this effect. We would like to
emphasize that for our method the registration of the EF radiation
caused by the ST effect plays the vital role, since only in this
case one can obtain true amplitude and phase characteristics of
the $E_x(H)$ dependence.

\begin{figure}
\includegraphics[width=8cm]{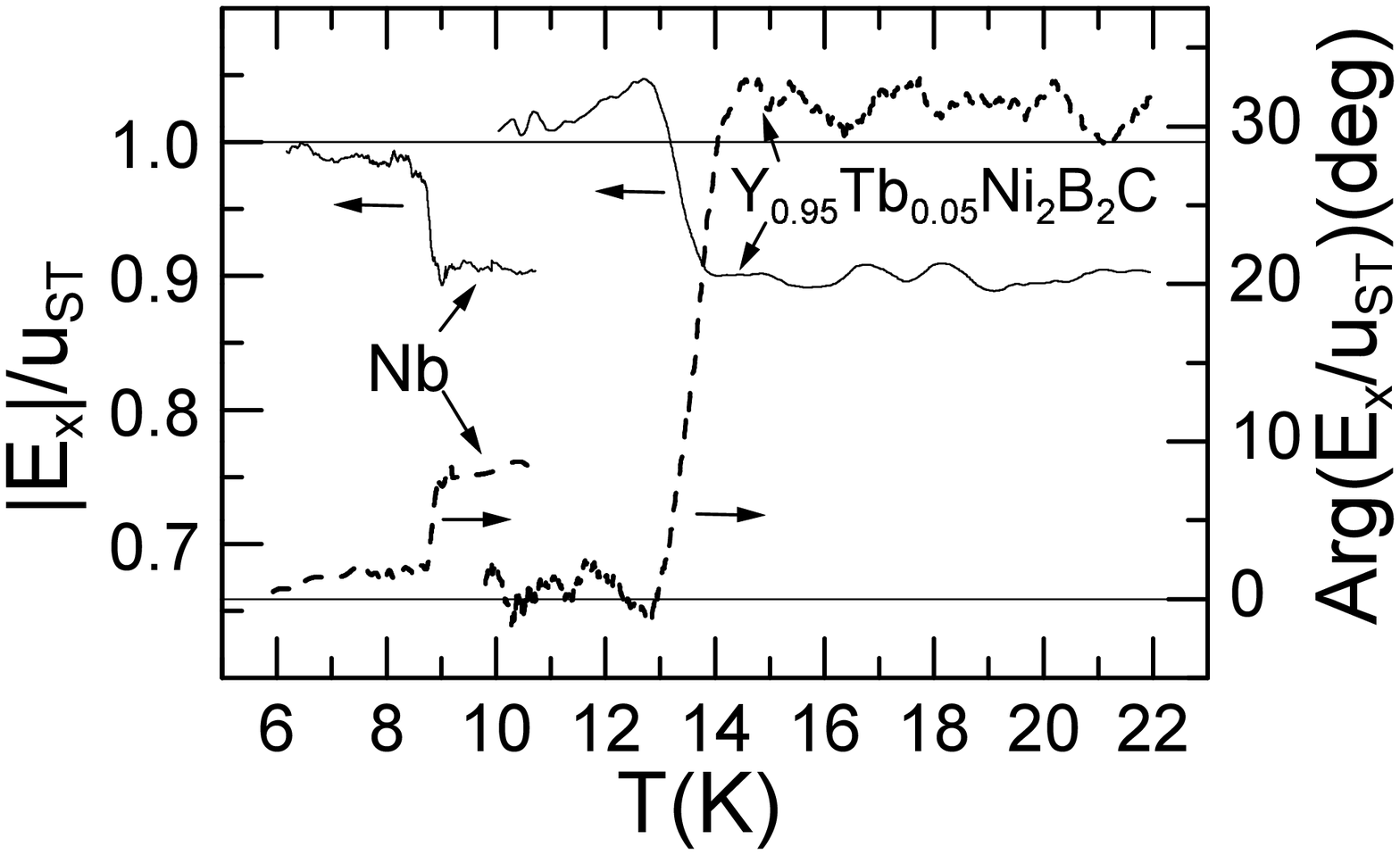}
 \caption{\label{f2} The modulus (solid lines) and the phase (dashed lines) of the EF caused by
the acoustic Stewart-Tolman effect ($H$=0)}
\end{figure}

Let us mention one technical detail important for the measurement
of $E_x$. When the receiving antenna is oriented along the weak
component $E_x$, the measured field $E_{meas}(H)$ contains the
admixture of the $E_y$ component caused by the edge effects:
$E_{meas}(H)=E_x(H)+\gamma E_y(H)$, where $\gamma$ is the complex
valued geometrical factor that remains nonzero at any orientation
of the antenna. The extraction of the $E_x(H)$ component can be
done using different parity of the $E_x(H)$ and $E_y(H)$
components with respect to $H$:
$E_x(H)=[E_{meas}(H)+E_{meas}(-H)]/2$. The examples of the real
(in-phase with $u_{ST}$) and the imaginary (quadrature) parts of
the $E_x$ field are shown in Fig.\,\ref{f3}.

\begin{figure}
\includegraphics[width=8cm]{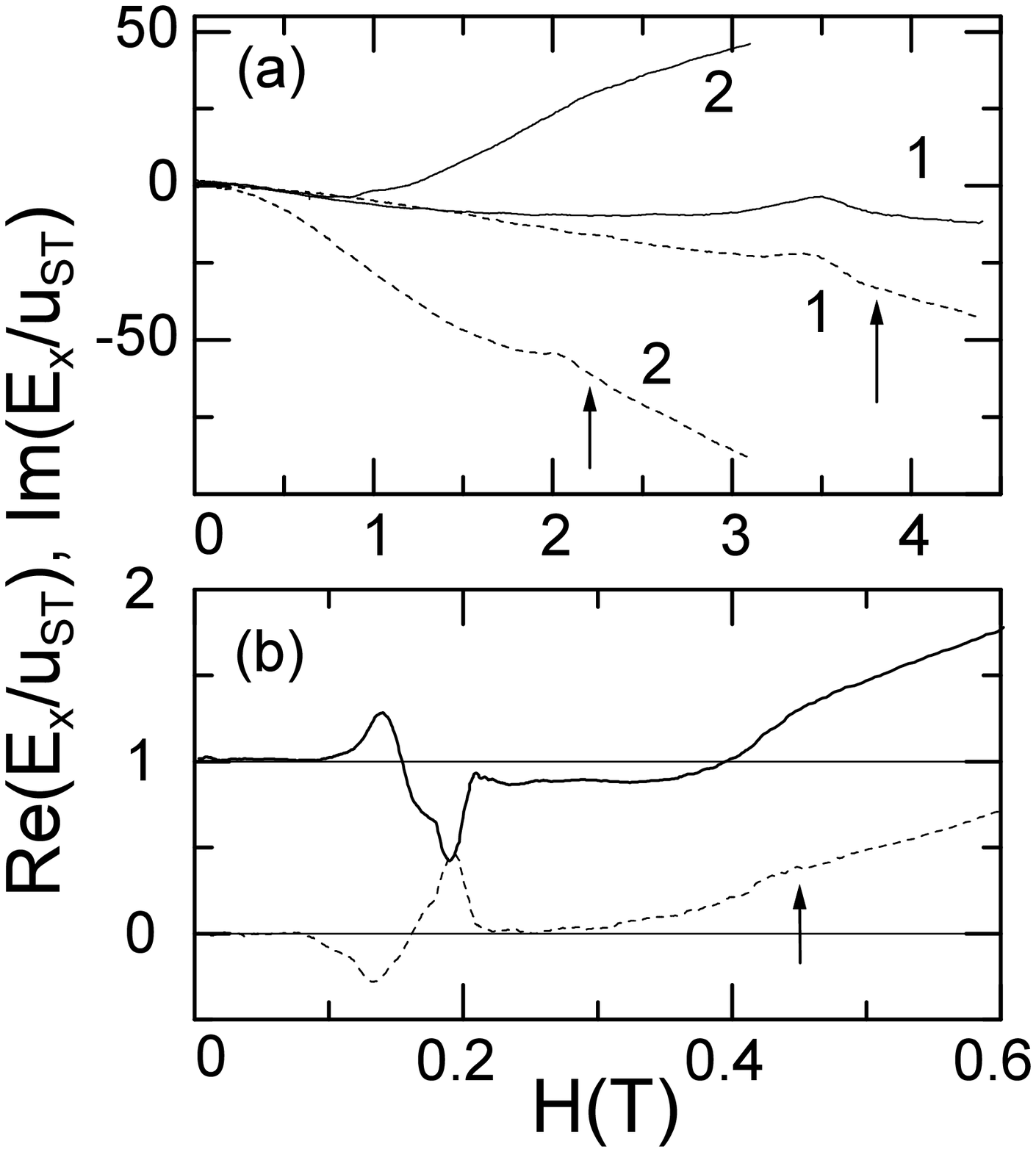}
 \caption{\label{f3} The real (solid lines) and the imaginary (dotted
 lines) parts of $E_x$ for the borocarbides (1 -
 Y$_{0.95}$Tb$_{0.05}$Ni$_2$B$_2$C at $T=1.7$ K, 2 - LuNi$_2$B$_2$C
 at $T=9$ K) (a) and for Nb at $T=1.7$ K (b). The arrows indicate
 $H_{c2}$.}
\end{figure}

One can easily find  that in the normal state the sign of the
imaginary part of the second term in the r.h.s. of Eq. (\ref{4a})
coincides  with the sign of $\Omega$. Thus, in borocarbides the
carriers are  of the electron type that is in agreement with the
transport data \cite{1}.  At the same time, in Nb the carriers are
the holes  (as well known, see. e.g. \cite{9}).

Eq. (\ref{4}) can be presented in the form
\begin{equation}\label{4b}
\frac{E_{x}(H)}{u_{ST}
X(H)}=1-i\beta_{dm}(H)-\frac{[1-X(H)]^2}{X(H)}
\left[\frac{\omega\alpha_M}{q^2 \frac{H^2}{4
\pi}}\right]\frac{|e|H}{m_e \omega c}.
\end{equation}
One can see that for real valued $\beta_{dm}$ and $\alpha_M$ the
function $\alpha_M(H)$ can be found from the real part of Eq.
(\ref{4b}) without the knowledge of the exact form of
$\beta_{dm}(H)$.

Eq. (\ref{4b}) is applicable for the normal state, as well, under
the replacements of the ratio $4\pi\omega\alpha_M/q^2 H^2$ in the
r.h.s. of Eq. (\ref{4b}) to the quantity $ |k_n^2| \Omega\tau$. To
present the data in a compact form it is convenient to define
formally the parameter $\omega \alpha_M$ at $H>H_{c2}$ as
$\omega\alpha_M = q^2 |k_n^2| H_{c2}^2 \Omega\tau/4\pi$.

The obtained dependencies of the Magnus coefficient on the
magnetic field  for the borocarbides are shown in Fig.\,\ref{f4}.
One can see from Fig.\,\ref{f4} that for the Y-based borocarbides
the dependence $\omega \alpha_M(H)$ is roughly linear in the whole
range of the magnetic field. At $H>H_{c2}$ one can expect the
linear increase of $\omega \alpha_M(H)$ just from the definition
of this quantity in the normal state (see above). At $H<H_{c2}$
the linear dependence is no more than the reflection of the fact
that the Magnus force is proportional to the number of vortices.
In LuNi$_2$B$_2$C the dependence $\omega \alpha_M(H)$ is almost
saturated in the normal state. The latter is connected with a
strong non-linearity of $\Omega \tau$, reported, for the first
time, in Ref. \cite{1}. We note that the coefficient $\alpha_M$
obtained is in two orders of magnitude smaller than $\eta$ (Ref.
\cite{8}) at the same $H$.

\begin{figure}
\includegraphics[width=8cm]{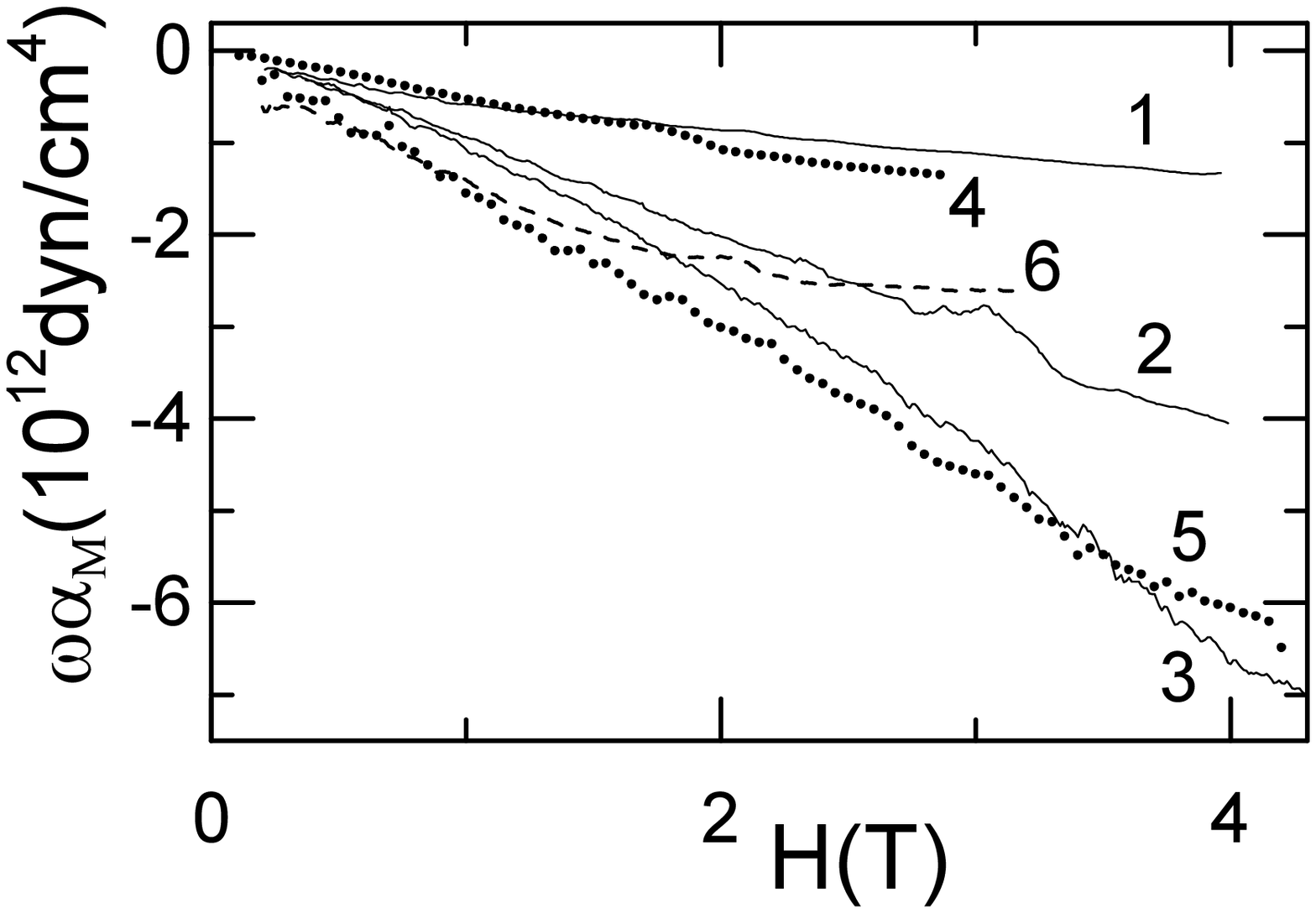}
 \caption{\label{f4} Field dependencies of the Magnus coefficient.
  Solid lines 1,2, and 3 -  Y$_{0.95}$Tb$_{0.05}$Ni$_2$B$_2$ at $T=8$ K ($H_{c2}=1.7$ T); 4 K (3.3 T); and 1.7 K (3.8
  T),
 correspondingly; dotted lines 4 and 5 -  YNi$_2$B$_2$C at 8 K (2 T) and 5 K (3.8 T), correspondingly;
 dashed line 6 - LuNi$_2$B$_2$C  at 9 K (2.15 T).}
\end{figure}

It is interesting to compare  the value of the  Hall coefficient
$R_H=\Omega\tau/(\sigma_0 H)$ that follows from our experimental
data with the results obtained by the transport measurements. For
the Y-based compounds at $H=4$ T and $T=5$ K we have $R_H\approx
2.8\cdot 10^{-12}$ $\Omega\cdot$cm$\cdot$Oe$^{-1}$ that almost
coincides with the results of Ref. \cite{1}. For LuNi$_2$B$_2$C at
$H=4$ T and $T=9$ K  we find $R_H\approx 5.3\cdot 10^{-12}$
$\Omega\cdot$cm$\cdot$Oe$^{-1}$ that is also close to the value
given in \cite{1}. For Nb we have $R_H\approx 6.5\cdot 10^{-13}$
$\Omega\cdot$cm$\cdot$Oe$^{-1}$ (compare with \cite{3n}).

The quantity $\overline{\omega \alpha_M}= \omega \alpha_M (4
\pi/q^2 |k_n^2| (H_{c2}^2)(\phi_0/B)$ ($\phi_0$ is the flux
quantum) is shown in Fig.\,\ref{f5}. For the mixed state these
data yield the Magnus coefficient for one vortex of unit length
(in units of $q^2 |k_n^2| H_{c2}^2/4\pi$).

\begin{figure}
\includegraphics[width=8cm]{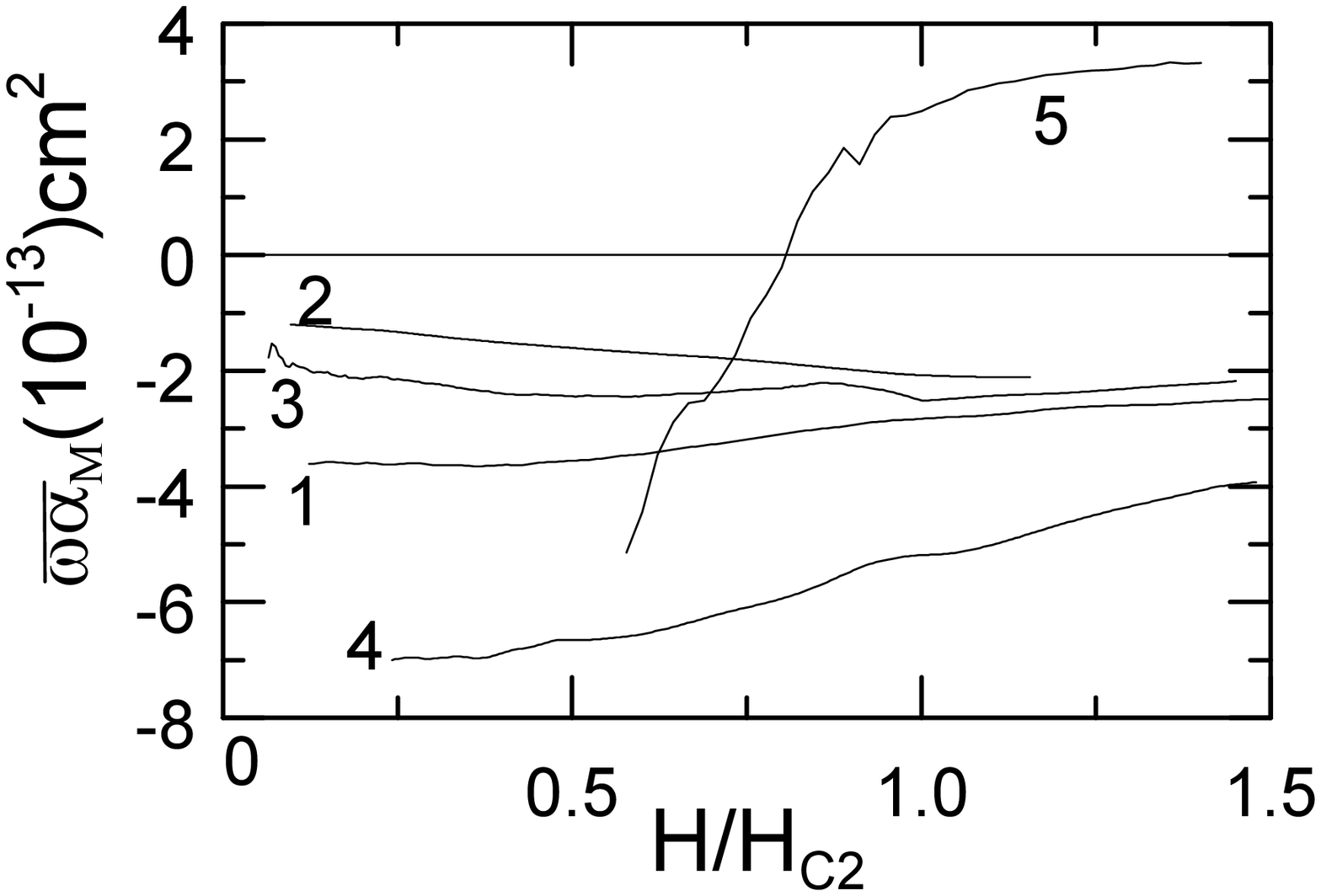}
 \caption{\label{f5} The Magnus coefficient for one vortex of unit length.
  1, 2 -  Y$_{0.95}$Tb$_{0.05}$Ni$_2$B$_2$
 at 8 K  and 1.7 K, correspondingly; 3 - YNi$_2$B$_2$C at 8 K; 4 - LuNi$_2$B$_2$C at 9 K; 5 - Nb at 1.7 K.}
\end{figure}

The qualitative difference in behavior of the gyroscopic force in
Nb is apparent already from Fig.\,\ref{f3}b. The in-phase with
$u_{ST}$ component of $E_x/u_{ST}$ becomes smaller than unity,
i.e. the gyroscopic term in $E_x$ changes its sign. In Nb the
parameter $\kappa$ is small and one cannot neglect the difference
between $B$ and $H$. To apply the procedure of finding of
$\alpha_M$ described above we should know the magnetic
permeability $\mu(H)=B/H$. To evaluate this function we use the
phenomenological expression for $\eta$: $i\omega\eta=k_n^2 (B
H_{c2}/4\pi)[1+\delta (1-B^2/H_{c2}^2)]$, where $\delta$ is the
fitting parameter. This expression is in better agreement with the
theoretical estimates \cite{2} and with the experiment \cite{3}
than the Bardin-Stephen formula. If the quantity $i\omega\eta$ is
given, the dependencies $\mu(H)$ and $\alpha_L(H)$ can be
determined from the experimental data. The result for $\mu(H)$
should be a smooth bounded function ($|E_y(H)|/H\leq\mu(H)\leq 1$,
where $E_y$ is  in units defined above). This requirement is
satisfied for $\delta\approx 0.6\div 0.7$. The further procedure
of the obtaining of $\alpha_M$ is the same as before. The result
for Nb is also presented in Fig. 5. The main qualitative
conclusion is that in Nb the sign of the Magnus force in the mixed
state is opposite to the sign of the gyroscopic force in the
normal state.

In conclusion, we have used the acoustic ST effect for realization
of the method of measurement of the Magnus force  in type II
superconductors that does not require the overthreshold flow of
vortices. The measurements in non-magnetic borocarbides with
conductivity of the electron type show that  the gyroscopic force
remains almost unchanged under the transition from the normal to
the mixed state. At the same time, in Nb this force changes its
sign below the transition point.

This study is supported in part by CRDF Foundation (Grant No
UP1-2566-KH-03) and by INTAS (Grant No 03-51-3036). We would like
to thank N.B.Kopnin for helpful discussions.

\end{document}